# Hysteretic Magnetoresistance in a Non-Magnetic SrSnO$_3$ Film via Thermal Coupling to Dynamic Substrate Behavior

Laxman Raju Thoutam[*,1], Tristan K. Truttmann, Anil Kumar Rajapitamahuni, and Bharat Jalan[*]

Department of Chemical Engineering and Materials Science

University of Minnesota - Twin Cities, Minneapolis, Minnesota 55455, USA

[1]Now at Department of Electronics and Communications Engineering, SR University, Warangal Urban, Telangana, 506371, India

[*] Corresponding authors: tlaxmanraju@gmail.com and bjalan@umn.edu

**Abstract**

Hysteretic magnetoresistance (MR) is often used as a signature of ferromagnetism in conducting oxide thin films and heterostructures. Here, magnetotransport is investigated in a non-magnetic uniformly La-doped $SrSnO_3$ film grown using hybrid molecular beam epitaxy. A 12 nm La:$SrSnO_3$/2 nm $SrSnO_3$/$GdScO_3$ (110) film with insulating behavior exhibited a robust hysteresis loop in the MR at $T < 5$ K accompanied by an anomaly at $\sim \pm 3$ T at $T < 2.5$ K. Furthermore, MR with the field in-plane yielded a value exceeded 100% at 1.8 K. Using detailed temperature-, angle- and magnetic field-dependent resistance measurements, we illustrate the origin of hysteresis is not due to magnetism in the film but rather is associated with the magnetocaloric effect of the $GdScO_3$ substrate. Given $GdScO_3$ and similar substrates are commonly used in complex oxide research, this work highlights the importance of thermal coupling to processes in the substrates which must be carefully accounted for in the data interpretation for thin films and heterostructures utilizing these substrates.

**Introduction:**

The choice of substrate has a dramatic effect on complex oxide film properties. Most obviously, the use of a chemically compatible and lattice-matched promotes high structural quality of the grown film. The complex oxide thin film community has also discovered more nuanced effects from the substrate including the ability to strain engineer films[1-4], confer its octahedral rotations to the films[5-7], and the use of an intrinsic electric field to electrostatically dope the film[8-11]. However, these effects are all static in the sense that the substrate is bestowing properties to the film through a fixed structural or physical relationship; there has been very little attention paid to how dynamic behavior in the substrate might influence the film. For instance, if a substrate exhibits the magnetocaloric effect (i.e., the generation or consumption of heat due to a change in applied magnetic field), it can potentially have a tremendous impact on the magnetotransport behavior of the films. Therefore, it can also hinder our ability to interpret data and may even lead to misinterpretation.

To study this effect, we use $SrSnO_3$ (SSO) thin films. This material belongs to family of non-magnetic alkaline-earth stannates ($ASnO_3$; $A$ = Ca, Sr or Ba), which have gained significant interest in recent years owing, in large part, to their (ultra)wide bandgap and high room-temperature electron mobility.[12, 13] These characteristics of stannates have paved the way for their use in transparent and high-power electronic applications. Motivated by this, significant progress has been made in synthesizing $BaSnO_3$ (BSO) and (SSO) films with atomic layer control using a variety of growth methods[14-20]. Although there are no studies of transport in bulk SSO single crystals, several exciting developments have occurred in SSO thin films including the demonstration of oxygen vacancy-induced room-temperature ferromagnetism[21, 22], high conductivity[23], weak localization and Aronov–Altshuler electron-electron interaction[24], a large phase coherence length[24] and the demonstration of several field-effect devices[25-27] and also at GHz frequencies[28]. As we will show, the insulating transport behavior achievable in this material make it a highly sensitive probe for the subtle thermal effects arising from the substrate. We therefore choose this material as a case study in thermal coupling between a film and processes that occur in the substrate. We note that our observation is applicable to any other films and heterostructures on these substrates with resistivity that is a strong function of temperature.

Using 12 nm La-doped SSO films grown on a 2 nm insulating SSO buffer layer on a $GdScO_3$ (GSO) (110) substrate, we discovered a large magnetoresistance (MR) at 1.8 K exceeding 100% accompanied by a hysteretic behavior. While this may *appear* to be indicative of exotic magnetism in otherwise non-magnetic SSO films, we show otherwise. Rather, it is due to thermal coupling between the film and the magnetocaloric behavior of the GSO substrate. We provide critical insight into the role of substrates on the magnetotransport measurements of films despite them being insulating.

The SSO/GSO film used in this study was grown using hybrid MBE.[29, 30] Figure 1(a) shows a high-resolution on-axis wide-angle X-ray diffraction 2θ–ω scan of a 12 nm La:SSO/2 nm SSO/GSO (110) revealing an expanded out-of-plane lattice parameter of 4.102 ± 0.002 Å. This value is consistent with the coherently strained tetragonal SSO phase on GSO substrate.[29] The SSO film is thus phase-pure, epitaxial, and fully strained on the GSO (110) substrate. Finite-thickness fringes were also observed, indicating laterally homogeneous films and an abrupt smooth interface. The streaky reflection high-energy electron diffraction (RHEED) pattern (Figure 1(b)) and atomic force microscopy (AFM) images (Figure 1(c)) confirm atomically smooth surfaces.

Figure 2 shows the magnetotransport results from this sample showing temperature-dependent sheet resistance ($R_s$) and MR. The room-temperature Hall-effect measurements measured a carrier concentration of $1.9 \times 10^{19}$ cm$^{-3}$ in this sample. Figure 2(a) shows an insulating-like behavior, i.e. an increasing sheet resistance with decreasing temperature, which we attribute to the disorder at the surface[31]. A sharp rise in R$_s$ was observed at $T < 6$ K. To investigate the low-temperature transport mechanism, we performed the Zabrodski analysis (see inset of Figure 2(a)) yielding a slope of 0.6. This result reveals that the low-temperature ($T < 6$ K) insulating behavior is associated with the Efros-Shklovskii (ES) variable range hopping. The exponential temperature dependence from this mechanism will be important for its strong response to small temperature changes. We also measured the MR, defined as $100\% \times [R(B) - R(0)] / R(0)$, where $R(B)$ is the sheet resistance at a magnetic-field, $B$. The MR with the field out of plane was measured at different temperatures by sweeping $B$ from 0 T → 9 T → -9 T → 0 T at 100 Oe/sec. Figures 2(b) and 2(c) show MR as a function of $B$ at 6 K $\leq T \leq$ 200 K and 1.8 K $\leq T \leq$ 5 K, respectively. The evolution of MR as a function of temperature at $B = 9$ T is shown in Figure 2(d). Several observations can be made here. First, with decreasing $T$, a crossover from a negative MR (6 K $\leq T \leq$ 200 K) to a combination of a negative and positive MR (1.8 K $\leq T \leq$ 5 K) occurs. Second, the magnetic field at which a crossover from negative to positive MR occurs decreases with decreasing temperature. Third, a surprising but robust hysteresis behavior appears at 1.8 K $\leq T \leq$ 5 K accompanied by an anomaly around $\pm 3$ T at $T \leq 2.4$ K (see inset of Figure 2(c)). We first discuss the origin of a negative and a positive MR and then return to the discussion of a rather unusual hysteresis behavior.

The presence of negative and positive MR in the hopping regime is seen previously in other semiconducting materials including n-type CuGaSe$_2$ [32], GaAs [33, 34] and most recently in stannates[35, 36]. The negative MR at low-field can be understood in terms of suppression of the destructive interference between forward scattering paths of the hopping electron for a given initial and final state of hop[34, 36]. Likewise, the positive MR can be explained by the shrinkage of the localized wave function of the electrons in presence of applied magnetic field resulting in the reduction of the overlap between neighboring hopping sites, causing an increase of resistance with increasing field[37]. The fact that the field at which crossover takes place decreases with decreasing temperature is again consistent with a crossover from a weak localization at higher temperatures to strongly localized transport at low temperatures.

To get further insights into the observed MR behavior, we performed angle-dependent MR measurements at 1.8 K using two different current directions ($I_y$ vs. $I_x$) as illustrated in the inset of Figures 3(a) and 3(b). We define the MR from these two configurations as $MR_1$ and $MR_2$. Regardless of the current directions ($I_y$ vs. $I_x$), both $MR_1$ and $MR_2$ revealed the suppression of low-field negative MR as field sweeps from the out-of-plane ($\theta_{xB} = 90°$) to the in-plane ($\theta_{xB} = 0°$) direction. This observation is again consistent with the previous reports ($In_2O_{3-x}$[38, 39], n-type GaAs[33]) of the angle-dependent anisotropic behavior in the strong localization regime. These results further revealed a strong anisotropy in the MR, specifically, in the high-field regime, as a function of $\theta_{xB}$. $MR_1$ and $MR_2$ reach ~100% and ~75% at 1.8 K, respectively when $\theta_{xB}$ reaches 0°, i.e. magnetic-field in the plane of the sample. The small difference in the values of the $MR_1$ and $MR_2$ is likely due to different starting values of the resistance, i.e. ($R_1(0) > R_2(0)$) in these two configurations. The observed high value of in-plane MR is again consistent with transport in the strong localization regime where the application of magnetic-field is expected to distort the electron wave function impacting the overlap between the hopping sites and consequently the MR[37]. A similar behavior was observed previously in GaAs in the strong localization regime[40, 41]. We further show in Figures 3(c) and 3(d) $MR_1$ and $MR_2$ measured at $\theta_{xB} = 90°$ and 0° as a function of temperature. Consistent with the onset of strong localization at $T < 6$ K, these results again revealed an increase in MR below ~ 6 K suggesting the origin of the large MR is associated with transport in the strong localization regime.

We now return to the discussion of the hysteresis and the anomaly in the MR measurements. For brevity, we only show $MR_1$, i.e. when current is along y-direction (see the inset of Figure 3(a)). Before we discuss this data, it is important to show the experimental setup illustrating the sample mount and how sample temperature was recorded. As shown in Figure 4(a), the sample is first glued onto a sample holder, which is then mounted onto the Dynacool™ rotator puck, which has an on-board thermocouple (shown as yellow dots in Figure 4(a)) that accurately measures the temperature of the sample. Figure 4(b) shows $MR_1$ at $\theta_{xB} = 90°$ and $T = 1.8$ K during the field sweep 0 T → 9 T → -9 T → 0 T at a rate of 100 Oe/sec reveling a clear hysteresis and an anomaly at ~ ± 3 T (marked as $B_c$). The anomaly occurs at an identical critical field, $B_c$ during the forward and reverse field sweeps. This can be seen clearly in Figure S2 using the first derivate of the $MR_1$ with respect to *B*. Figure 4(c) shows the thermocouple temperature deviation from the setpoint during the sweep as a function of *B*. This plot reveals a hysteretic temperature behavior and a sudden change in temperature at $B_c$ – the same critical field where an anomaly in MR was seen. The change in temperature at $B_c$ is ~ 8 mK. We attribute the hysteretic temperature and the sudden change in temperature at $B_c$ being responsible for the MR hysteresis and the anomaly. Since the low-temperature resistance of the sample is an exponential function of temperature, even small changes in temperature can introduce observable changes in the resistance. However, it raises additional questions: what is the origin of hysteretic temperature, and why there is a sudden change in temperature at $B_c$?

To investigate these questions, we measured the variation in temperature of a bare GSO (110) substrate without any SSO film in the same way as in Figure 4(c). Surprisingly, the bare substrate yielded an identical hysteretic behavior in temperature as a function of magnetic field accompanied by the same anomaly at $B_c$ (Figure S2). The bare substrate also showed anomaly in temperature-hysteresis at $B_c$, which persisted at $T \leq 2.4$ K. The value of $B_c$ from the temperature-hysteresis of the bare GSO substrate, and from the MR-hysteresis of the 12 nm La:SSO/2 nm SSO/GSO (110) is depicted in Figure 4(d) and 4(e) as a function of $T$ and $\theta_{xB}$, respectively. Open symbols correspond to data from the La:SSO/SSO/GSO whereas the solid symbols represent the bare GSO substrate. An identical value of $B_c$ as a function of $T$ and $\theta_{xB}$ was observed between La:SSO/SSO/GSO and the bare substrate. The difference between $B_c$ for the out-of-plane and in-plane directions is likely due to the distorted orthorhombic structure of GSO substrate. A similar magnetic anisotropy with a preferred easy axis and hard axis along different crystallographic directions was seen in a similar rare-earth perovskite scandate substrate $DyScO_3$ [42]. To this end, it becomes apparent that the origin of hysteresis and the anomaly in MR is related to the substrate.

But then why does the anomaly disappear at $T > 2.4$ K and what is the physical significance of $B_c$? We explain this due to the antiferromagnetic transition in GSO. GSO has a distorted orthorhombic structure with an antiferromagnetic ordering below the Néel temperature $T_N = \sim 2.6$ K[43], and is also shown to exhibit reversible magnetocaloric properties[44]. An external magnetic field disfavors the antiferromagnetic state, so below $T_N$, a magnetic field can induce a phase transition to the paramagnetic state. Thus, the feature associated with the anomaly in the MR at $T < 2.4$ K can be attributed to the magnetic-field induced phase-transition between GSO's anti-ferromagnetic at low fields to its paramagnetic state at high fields. In the paramagnetic state, increasing magnitude of applied field aligns the spin dipoles to field direction. This spin alignment lowers the enthalpy of the spin system, which is balanced by releasing heat to the lattice. This behavior raises the sample temperature and consequently lowers the film resistance resulting in the hysteresis above the critical field or for any field for temperatures immediately above $T_N$. According to this description, decreasing ramp rates should lower hysteresis, which we confirmed in Figure S3 and the accompanying discussion.

In summary, using a detailed magnetotransport measurements in La-doped SSO films on GSO substrate, we discuss the origin of a large MR exceeding 100% in an insulating SSO film owing to the transport in the strong localization regime. It was found that this sample also exhibits a robust hysteresis in MR suggesting an apparent emergence of exotic magnetism in an otherwise non-magnetic La-doped SSO film. However, we provide clear evidence that thermal coupling between SSO film and the magnetocaloric effect of the GSO substrate is responsible for the hysteresis behavior. We further found an anomaly in MR which was attributed to the field-induced antiferromagnetic → paramagnetic phase transition in GSO substrate below the Néel temperature. Two other commonly used rare-earth scandate substrates $TbScO_3$ and $DyScO_3$ have also demonstrated antiferromagnetism with similar Néel temperatures[42, 43, 45], and a strong magnetocaloric effect has also been measured for $DyScO_3$[46]. These findings further suggest that the thermal coupling between films and substrates — especially these rare-earth scandate substrates — must be considered in the interpretation of any ostensible magnetism observed in complex oxides heterostructures.

**Acknowledgement:**

The authors acknowledge Fengdeng Liu for the illustration of the Dynacool™ rotator puck. This work was supported primarily by the National Science Foundation (NSF) through the University of Minnesota MRSEC under Award Number DMR-2011401. Part of this work was supported through the Air Force

Office of Scientific Research (AFOSR) through Grant No. FA9550-19-1-0245 and FA9550-21-1-0025 and through NSF DMR-1741801. Portions of this work were conducted in the Minnesota Nano Center, which is supported by the National Science Foundation (NSF) through the National Nano Coordinated Infrastructure Network (NNCI) under Award Number ECCS-1542202. Part of this work was also carried out in the College of Science and Engineering Characterization Facility, University of Minnesota, which has received capital equipment funding from the NSF through the UMN MRSEC program.

**Data Availability**

The data that support the findings of this study are available from the corresponding author upon reasonable request.

**Experiment**

The epitaxial La-doped SSO film was grown on GSO (110) substrates using a radical-based molecular beam epitaxy approach. Details of growth condition can be found elsewhere[18, 29, 30, 36]. A brief discussion is presented here. Sr and La were evaporated using effusion cells, and oxygen was supplied using RF plasma source operating at 250 W. Sn is supplied using a metal-organic chemical precursor, hexamethylditin (HMDT). The film was grown at a substrate temperature of 950 °C (thermocouple temperature), and at a fixed oxygen background pressure of $5 \times 10^{-6}$ Torr. Electronic transport measurements were performed in a Quantum Design Physical Property Measurement System (Quantum Design Dynacool™) using a van der Pauw configuration. Indium was used as an Ohmic contact.

Room-temperature electron mobility in thin La-doped BSO films has now reached a value of 183 $cm^2V^{-1}s^{-1}$.[15] Although, the mobility in BSO is yet to match that of bulk single crystals (320 $cm^2V^{-1}s^{-1}$)[47], thin films have facilitated several fundamental studies of structure-defect-property relationships including the investigation of film/substrate lattice mismatch[14, 19, 48-50], oxygen vacancies[16, 51-53], charged defects[14, 49, 52, 54] and non-stoichiometry[14, 36, 55] on electronic transport. Several device structures have also been demonstrated.[56-62] Similarly, the room-temperature mobility in La-doped SSO films has now reached 70 $cm^2V^{-1}s^{-1}$ in a film as thin as 12 nm[18]. In comparison to BSO, SSO offers a smaller lattice parameter (4.034 Å vs 4.116 Å) and a higher bandgap (~4 eV vs ~3 eV).[29, 63]

**References:**

1. Ahadi, K.; Galletti, L.; Li, Y.; Salmani-Rezaie, S.; Wu, W.; Stemmer, S., Enhancing superconductivity in SrTiO$_3$ films with strain. *Science Advances* **2019,** *5* (4), eaaw0120.
2. Hu, S.; Cazorla, C.; Xiang, F.; Ma, H.; Wang, J.; Wang, J.; Wang, X.; Ulrich, C.; Chen, L.; Seidel, J., Strain Control of Giant Magnetic Anisotropy in Metallic Perovskite SrCoO$_{3-\delta}$ Thin Films. *ACS Applied Materials & Interfaces* **2018,** *10* (26), 22348-22355.
3. Schlom, D. G.; Chen, L.-Q.; Eom, C.-B.; Rabe, K. M.; Streiffer, S. K.; Triscone, J.-M., Strain Tuning of Ferroelectric Thin Films. *Annual Review of Materials Research* **2007,** *37* (1), 589-626.
4. Schlom, D. G.; Chen, L.-Q.; Fennie, C. J.; Gopalan, V.; Muller, D. A.; Pan, X.; Ramesh, R.; Uecker, R., Elastic strain engineering of ferroic oxides. *MRS Bulletin* **2014,** *39*, 118.
5. Rondinelli, J. M.; May, S. J.; Freeland, J. W., Control of octahedral connectivity in perovskite oxide heterostructures: An emerging route to multifunctional materials discovery. *MRS Bulletin* **2012,** *37*, 261.
6. Liao, Z.; Huijben, M.; Zhong, Z.; Gauquelin, N.; Macke, S.; Green, R. J.; Van Aert, S.; Verbeeck, J.; Van Tendeloo, G.; Held, K.; Sawatzky, G. A.; Koster, G.; Rijnders, G., Controlled lateral anisotropy in correlated manganite heterostructures by interface-engineered oxygen octahedral coupling. *Nature Materials* **2016,** *15*, 425.
7. Bousquet, E.; Dawber, M.; Stucki, N.; Lichtensteiger, C.; Hermet, P.; Gariglio, S.; Triscone, J.-M.; Ghosez, P., Improper ferroelectricity in perovskite oxide artificial superlattices. *Nature* **2008,** *452*, 732.
8. A. Ohtomo; Hwang, H. Y., A high-mobility electron gas at the LaAlO$_3$/SrTiO$_3$ heterointerface. *Nature* **2004,** *427*, 423.
9. Kim, U.; Park, C.; Ha, T.; Kim, Y. M.; Kim, N.; Ju, C.; Park, J.; Yu, J.; Kim, J. H.; Char, K., All-perovskite transparent high mobility field effect using epitaxial BaSnO$_3$ and LaInO$_3$ featured. *APL Mater.* **2015,** *3*, 036101.
10. Liu, C.; Yan, X.; Jin, D.; Ma, Y.; Hsiao, H.-W.; Lin, Y.; Bretz-Sullivan, T. M.; Zhou, X.; Pearson, J.; Fisher, B.; Jiang, J. S.; Han, W.; Zuo, J.-M.; Wen, J.; Fong, D. D.; Sun, J.; Zhou, H.; Bhattacharya, A., Two-dimensional superconductivity and anisotropic transport at KTaO$_3$ (111) interfaces. *Science* **2021,** *371*, 716.
11. P. Xu, Y. A., C. Cheng, V. S. Pribiag, R. B. Comes, P. V. Sushko, S. A. Chambers, and B. Jalan, Predictive control over charge density in the two-dimensional electron gas at the polar/non-polar NdTiO$_3$/SrTiO$_3$ interface. *Phys. Rev. Lett.* **2016,** *117*, 106803.

12. Lee, W.-J.; Kim, H. J.; Kang, J.; Jang, D. H.; Kim, T. H.; Lee, J. H.; Kim, K. H., Transparent Perovskite Barium Stannate with High Electron Mobility and Thermal Stability. *Ann. Rev. Mater. Res.* **2017,** *47*, 391.
13. Prakash, A.; Jalan, B., Wide Bandgap Perovskite Oxides with High Room-Temperature Electron Mobility. *Adv. Mater. Interfaces* **2019,** *6*, 1900479.
14. A. Prakash; P. Xu; A. Faghaninia; S. Shukla; J. W. Ager; C. S. Lo; B. Jalan, Wide bandgap $BaSnO_3$ films with room temperature conductivity exceeding $10^4$ S cm$^{-1}$. *Nat. Comm.* **2017,** *8* (1), 15167.
15. Paik, H.; Chen, Z.; Lochocki, E.; H., A. S.; Verma, A.; Tanen, N.; Park, J.; Uchida, M.; Shang, S.; Zhou, B.-C.; Brützam, M.; Uecker, R.; Liu, Z.-K.; Jena, D.; Shen, K. M.; Muller, D. A.; Schlom, D. G., Adsorption-Controlled Growth of La-doped $BaSnO_3$ by molecular-Beam Epitaxy. *APL Mater.* **2017,** *5* (11), 116107.
16. Cho, H. J.; Feng, B.; Onozato, T.; Wei, M.; Sanchela, A. V.; Ikuhara, Y.; Ohta, H., Investigation of electrical and thermal transport property reductions in La-doped $BaSnO_3$ films. *Physical Review Materials* **2019,** *3* (9), 094601.
17. Ganguly, K.; Prakash, A.; Jalan, B.; Leighton, C., Mobility-electron density relation probed via controlled oxygen vacancy doping in epitaxial BaSnO3. *APL Mater.* **2017,** *5* (5), 056102.
18. Truttmann, T.; Prakash, A.; Yue, J.; Mates, T. E.; Jalan, B., Dopant solubility and charge compensation in La-doped $SrSnO_3$ films. *Applied Physics Letters* **2019,** *115* (15), 152103.
19. Raghavan, S.; Schumann, T.; Kim, H.; Zhang, J. Y.; Cain, T. A.; Stemmer, S., High-mobility $BaSnO_3$ grown by oxide molecular beam epitaxy. *APL Materials* **2016,** *4* (1), 016106.
20. Kim, H. J.; Kim, U.; Kim, T. H.; Kim, J.; Kim, H. M.; Jeon, B.-G.; Lee, W.-J.; Mun, H. S.; Hong, K. T.; Yu, J., Physical properties of transparent perovskite oxides (Ba, La) SnO 3 with high electrical mobility at room temperature. *Phys. Rev. B* **2012,** *86* (16), 165205.
21. Gao, Q.; Chen, H.; Li, K.; Liu, Q., Band Gap Engineering and Room-Temperature Ferromagnetism by Oxygen Vacancies in $SrSnO_3$ Epitaxial Films. *ACS Applied Materials & Interfaces* **2018,** *10* (32), 27503-27509.
22. Gao, D.; Gao, X.; Wu, Y.; Zhang, T.; Yang, J.; Li, X., Co-doped $SrSnO_3$ epitaxial thin films on MgO with tunable band gap and room-temperature ferromagnetism. *Physica E: Low-dimensional Systems and Nanostructures* **2019,** *109*, 101-106.
23. Wei, M.; Sanchela, A. V.; Feng, B.; Ikuhara, Y.; Cho, H. J.; Ohta, H., High electrical conducting deep-ultraviolet-transparent oxide semiconductor La-doped $SrSnO_3$ exceeding similar to 3000 Scm$^{-1}$. *Appl. Phys. Lett.* **2020,** *116*, 022103.
24. Yue, J.; Thoutam, L. R.; Prakash, A.; Wang, T.; Jalan, B., Unraveling the effect of electron-electron interaction on electronic transport in La-doped $SrSnO_3$ films. *Applied Physics Letters* **2019,** *115* (8), 082102.
25. Chaganti, V. R. S. K.; Prakash, A.; Yue, J.; Jalan, B.; Koester, S. J., Demonstration of a Depletion-Mode $SrSnO_3$ n-Channel MESFET. *IEEE Electron Device Letters* **2018,** *39* (9), 1381-1384.
26. L. R. Thoutam, J. Y., A. Prakash, T. Wang, K. E. Elangovan, and B. Jalan, Electrostatic control of insulator–metal transition in La-doped $SrSnO_3$ films. *ACS Appl. Mater. Interfaces* **2019,** *11*, 7666.
27. V. R. S. K. Chaganti, T. K. T., F. Liu, B. Jalan, S. J. Koester, $SrSnO_3$ field-effect transistors with recessed gate electrodes. *IEEE Electron Dev. Lett.* **2020,** *41*, 1428.
28. J. Wen, V. R. S. K. C., T. K. Truttmann, F. Liu, B. Jalan, and S. J. Koester, $SrSnO_3$ Metal-Semiconductor Field-Effect Transistor with GHz Operation. *IEEE Electron Dev. Lett.* **2021,** *42*, 74.
29. Wang, T.; Prakash, A.; Dong, Y.; Truttmann, T.; Bucsek, A.; James, R.; Fong, D. D.; Kim, J.-W.; Ryan, P. J.; Zhou, H.; Birol, T.; Jalan, B., Engineering $SrSnO_3$ Phases and Electron Mobility at Room Temperature Using Epitaxial Strain. *ACS Applied Materials & Interfaces* **2018,** *10* (50), 43802-43808.
30. A. Prakash, T. W., A. Bucsek, T. K. Truttmann, A. Fali, M. Cotrufo, H. Yun, J.-Woo Kim, P. J. Ryan, K. Andre Mkhoyan, A. Alù, Y. Abate, R. D. James, and B. Jalan, Self-Assembled Periodic Nanostructures Using Martensitic Phase Transformations. *Nano Lett.* **2021,** *21*, 1246.
31. T. K. Truttmann, F. L., J. G. Barriocanal, R. D. James, and B. Jalan, Strain Relaxation via Phase Transformation in $SrSnO_3$. *ACS Appl. Electron. Mater.* **2021,** *3*, 1127.

32. Lisunov, K. G.; Arushanov, E.; Thomas, G. A.; Bucher, E.; Schön, J. H., Variable-range hopping conductivity and magnetoresistance in n-CuGaSe$_2$. *Journal of Applied Physics* **2000,** *88* (7), 4128-4134.
33. Raikh, M. E.; Czingon, J.; Ye, Q.-y.; Koch, F.; Schoepe, W.; Ploog, K., Mechanisms of magnetoresistance in variable-range-hopping transport for two-dimensional electron systems. *Physical Review B* **1992,** *45* (11), 6015-6022.
34. Raikh, M. E.; Wessels, G. F., Single-scattering-path approach to the negative magnetoresistance in the variable-range-hopping regime for two-dimensional electron systems. *Physical Review B* **1993,** *47* (23), 15609-15621.
35. Wang, H.; Walter, J.; Ganguly, K.; Yu, B.; Yu, G.; Zhang, Z.; Zhou, H.; Fu, H.; Greven, M.; Leighton, C., Wide-voltage-window reversible control of electronic transport in electrolyte-gated epitaxial BaSnO$_3$. *Physical Review Materials* **2019,** *3* (7), 075001.
36. Wang, T.; Thoutam, L. R.; Prakash, A.; Nunn, W.; Haugstad, G.; Jalan, B., Defect-driven localization crossovers in MBE-grown La-doped SrSnO$_3$ films. *Physical Review Materials* **2017,** *1* (6), 061601.
37. Efros, B. I. S. a. A. L., *Electronic Properties of Doped Semiconductors*. Springer: Berling, 1984.
38. Faran, O.; Ovadyahu, Z., Magnetoconductance in the variable-range-hopping regime due to a quantum-interference mechanism. *Physical Review B* **1988,** *38* (8), 5457-5465.
39. Ovadyahu, Z., Anisotropic magnetoresistance in a Fermi glass. *Physical Review B* **1986,** *33* (9), 6552-6554.
40. Fischbach, J. U.; Rühle, W.; Bimberg, D.; Bauser, E., Experimental determination of the anisotropy of the exciton wave function of GaAs in a magnetic field. *Solid State Communications* **1976,** *18* (9), 1255-1258.
41. Qiu-yi Ye, B. I. S., A. Zrenner, F. Koch, and K. Poog, Hopping Transport in δ doping layers in GaAs *Phys. Rev . B* **1990,** *41*, 8477.
42. Ke, X.; Adamo, C.; Schlom, D. G.; Bernhagen, M.; Uecker, R.; Schiffer, P., Low temperature magnetism in the perovskite substrate DyScO$_3$. *Applied Physics Letters* **2009,** *94* (15), 152503.
43. Raekers, M.; Kuepper, K.; Bartkowski, S.; Prinz, M.; Postnikov, A. V.; Potzger, K.; Zhou, S.; Arulraj, A.; Stüßer, N.; Uecker, R.; Yang, W. L.; Neumann, M., Electronic and magnetic structure of RScO$_3$ (R = Sm, Gd, Dy)from x-ray spectroscopies and first-principles calculations. *Physical Review B* **2009,** *79* (12), 125114.
44. Wu, Y.-D.; Chen, H.; Hua, J.-Y.; Qin, Y.-L.; Ma, X.-H.; Wei, Y.-Y.; Zi, Z.-F., Giant reversible magnetocaloric effect in orthorhombic GdScO3. *Ceramics International* **2019,** *45* (10), 13094-13098.
45. Kamba, S.; Goian, V.; Nuzhnyy, D.; Bovtun, V.; Kempa, M.; Prokleška, J.; Bernhagen, M.; Uecker, R.; Schlom, D. G., Polar phonon anomalies in single-crystalline TbScO$_3$. *Phase Transitions* **2012,** *86*, 206.
46. Yao-DongWu; Qin, Y.-L.; Ma, X.-H.; Li, R.-W.; Wei, Y.-Y.; Zi, Z.-F., Large rotating magnetocaloric effect at low magnetic fields in the Ising-like antiferromagnet DyScO$_3$ single crystal. *J. Alloy Compd.* **2019,** *777*, 673.
47. Kim, H. J.; Kim, U.; Kim, H. M.; Kim, T. H.; Mun, H. S.; Jeon, B.-G.; Hong, K. T.; Lee, W.-J.; Chanjong, J.; Kim, K. H.; Char, K., High Mobility in a Stable Transparent Perovskite Oxide. *Appl. Phys. Express* **2012,** *5* (6), 061102.
48. Wadekar, P. V.; Alaria, J.; O'Sullivan, M.; Flack, N. L. O.; Manning, T. D.; Phillips, L. J.; Durose, K.; Lozano, O.; Lucas, S.; Claridge, J. B.; Rosseinsky, M. J., Improved electrical mobility in highly epitaxial La:BaSnO$_3$ films on SmScO$_3$(110) substrates. *Applied Physics Letters* **2014,** *105* (5), 052104.
49. Paik, H.; Chen, Z.; Lochocki, E.; Seidner H, A.; Verma, A.; Tanen, N.; Park, J.; Uchida, M.; Shang, S.; Zhou, B.-C.; Brützam, M.; Uecker, R.; Liu, Z.-K.; Jena, D.; Shen, K. M.; Muller, D. A.; Schlom, D. G., Adsorption-controlled growth of La-doped BaSnO$_3$ by molecular-beam epitaxy. *APL Materials* **2017,** *5* (11), 116107.

50. H. Yun, K. G., W. Postiglione, B. Jalan, C. Leighton, K. A. Mkhoyan, and J. S. Jeong,, Microstructure characterization of BaSnO3 thin films on LaAlO$_3$ and PrScO$_3$ substrates from transmission electron microscopy. *Sci. Rep.* **2018,** *8*, 10245.
51. Yoon, D.; Yu, S.; Son, J., Oxygen vacancy-assisted recovery process for increasing electron mobility in n-type BaSnO$_3$ epitaxial thin films. *NPG Asia Materials* **2018,** *10* (4), 363-371.
52. Ganguly, K.; Prakash, A.; Jalan, B.; Leighton, C., Mobility-electron density relation probed via controlled oxygen vacancy doping in epitaxial BaSnO$_3$. *APL Materials* **2017,** *5* (5), 056102.
53. Jaim, H.; Lee, S.; Zhang, X.; Takeuchi, I., Stability of the oxygen vacancy induced conductivity in BaSnO3 thin films on SrTiO$_3$. *Appl. Phys. Lett.* **2017,** *111*, 172102.
54. H. Yun, A. P., T. Birol, B. Jalan, and K. A. Mkhoyan, Direct observation and consequences of dopant segregation inside and outside dislocation cores in perovskite BaSnO$_3$. *Nano Lett.* **2021,** *21*, 4357.
55. Prakash, A.; Xu, P.; Wu, X.; Haugstad, G.; Wang, X.; Jalan, B., Adsorption-controlled growth and the influence of stoichiometry on electronic transport in hybrid molecular beam epitaxy-grown BaSnO3 films. *J. Mater. Chem. C* **2017,** *5* (23), 5730-5736.
56. Yue, J.; Prakash, A.; Robbins, M. C.; Koester, S. J.; Jalan, B., Depletion Mode MOSFET Using La-Doped BaSnO3 as a Channel Material. *ACS Appl. Mater. Interfaces* **2018,** *10* (25), 21061-21065.
57. Kim, U.; Park, C.; Ha, T.; Kim, Y. M.; Kim, N.; Ju, C.; Park, J.; Yu, J.; Kim, J. H.; Char, K., All-perovskite transparent high mobility field effect using epitaxial BaSnO3 and LaInO3. *APL Mater.* **2015,** *3* (3), 036101.
58. Kim, Y. M.; Park, C.; Ha, T.; Kim, U.; Kim, N.; Shin, J.; Kim, Y.; Yu, J.; Kim, J. H.; Char, K., High-k perovskite gate oxide BaHfO3. *APL Mater.* **2017,** *5* (1), 016104.
59. Park, C.; Kim, U.; Ju, C. J.; Park, J. S.; Kim, Y. M.; Char, K., High mobility field effect transistor based on BaSnO3 with Al2O3 gate oxide. *Appl. Phys. Lett.* **2014,** *105* (20), 203503.
60. Kim, Y. M.; Park, C.; Kim, U.; Ju, C.; Char, K., High-mobility BaSnO3 thin-film transistor with HfO2 gate insulator. *Appl. Phys. Express* **2015,** *9* (1), 011201.
61. Park, J.; Paik, H.; Nomoto, K.; Lee, K.; Park, B.; Grisafe, B.; Wang, L.; Salahuddin, S.; Datta, S.; Kim, Y.; Jena, D.; Xing, H.; Schlom, D. G., Fully transparent field-effect transistor with high drain current and on-off ratio. *APL Mater.* **2020,** *8*, 011110.
62. Chandrasekar, H.; Cheng, J.; Wang, T.; Xia, Z.; Combs, N. G.; Freeze, C. R.; Marshall, P. B.; McGlone, J.; Arehart, A.; Ringel, S.; Janotti, A.; Stemmer, S.; Lu, W.; Rajan, S., Velocity saturation in La-doped BaSnO$_3$ thin films. *Appl. Phys. Lett.* **2019,** *115*, 092102.
63. A. Prakash, N. F. Q., H. Yun, J. Held, T. Wang, T. Truttmann, J. M. Ablett, C. Weiland, T-L. Lee, J. C. Woicik, KA. Mkhoyan and B. Jalan, , Separating electrons and donors in BaSnO$_3$ via band engineering. *Nano Lett.* **2019,** *19*, 8920.

**Figures (Color Online):**

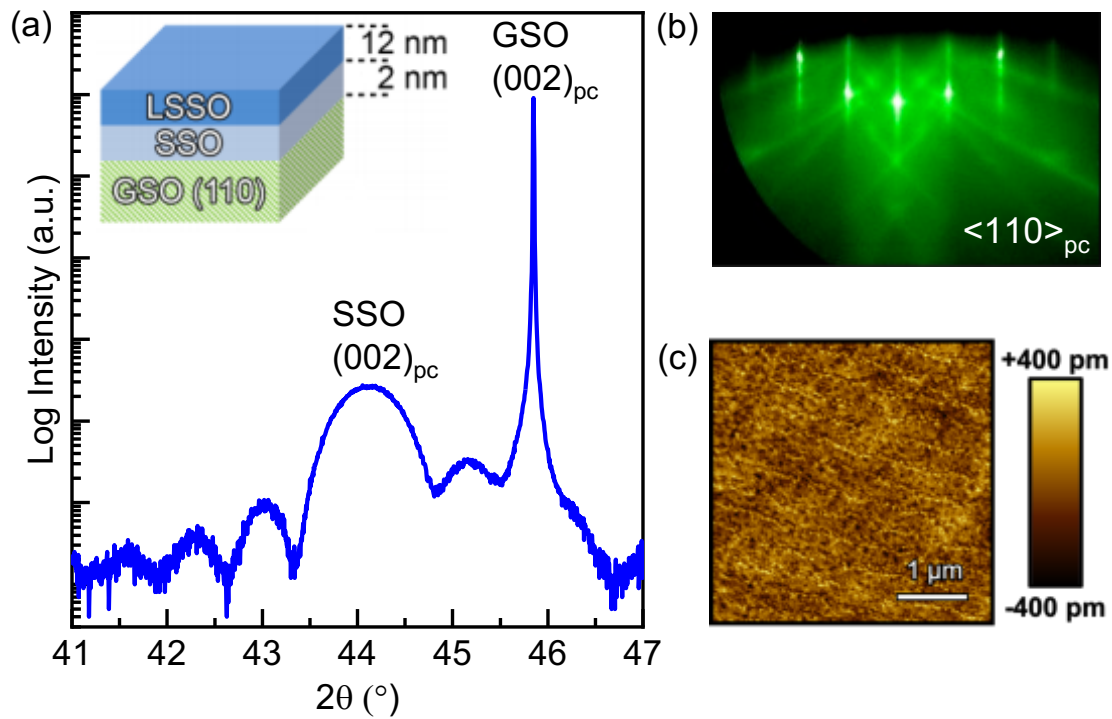

**Figure 1**: (a) High-resolution X-ray 2θ-ω coupled scan of a 12 nm La:SSO/2 nm SSO/GSO (110). The subscript pc denotes pseudocubic notation. The schematic of the structure is shown in the inset. (b) RHEED pattern along the substrate edge, and (c) atomic force microscopy of the sample, showing a smooth surface morphology.

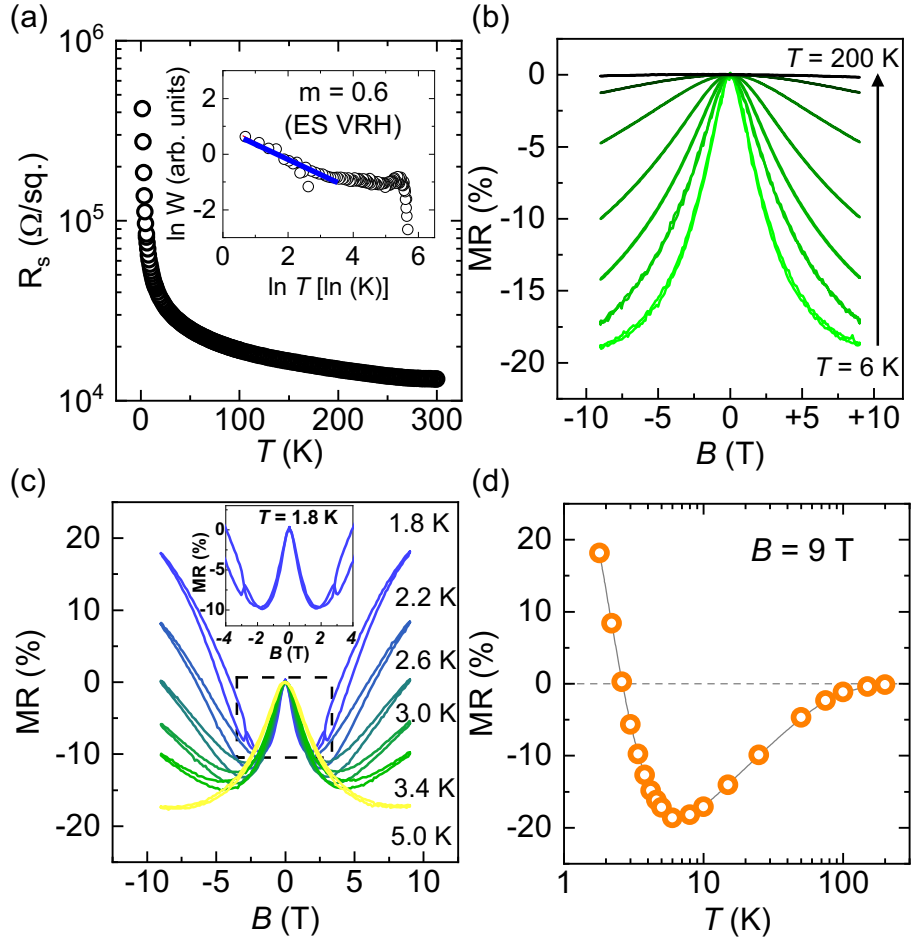

**Figure 2:** (a) Temperature-dependent sheet resistance ($R_s$) of 12 nm La:SSO/2 nm SSO/GSO (110). The inset shows the Zabrodskii analysis of the sample revealing insulating behavior at low temperatures, consistent with Efros-Shklovskii variable range hopping. (b) Perpendicular MR at various constant temperatures from 6 K $\leq T \leq$ 200 K. (c) Low temperature ($T \leq 5$ K) perpendicular MR shows a crossover from negative to positive magnetoresistance with applied field. Inset shows the zoom-in figure revealing an anomaly (sudden drop in resistance) and the hysteresis in MR. (d) MR as a function of temperature for $B = 9$ T.

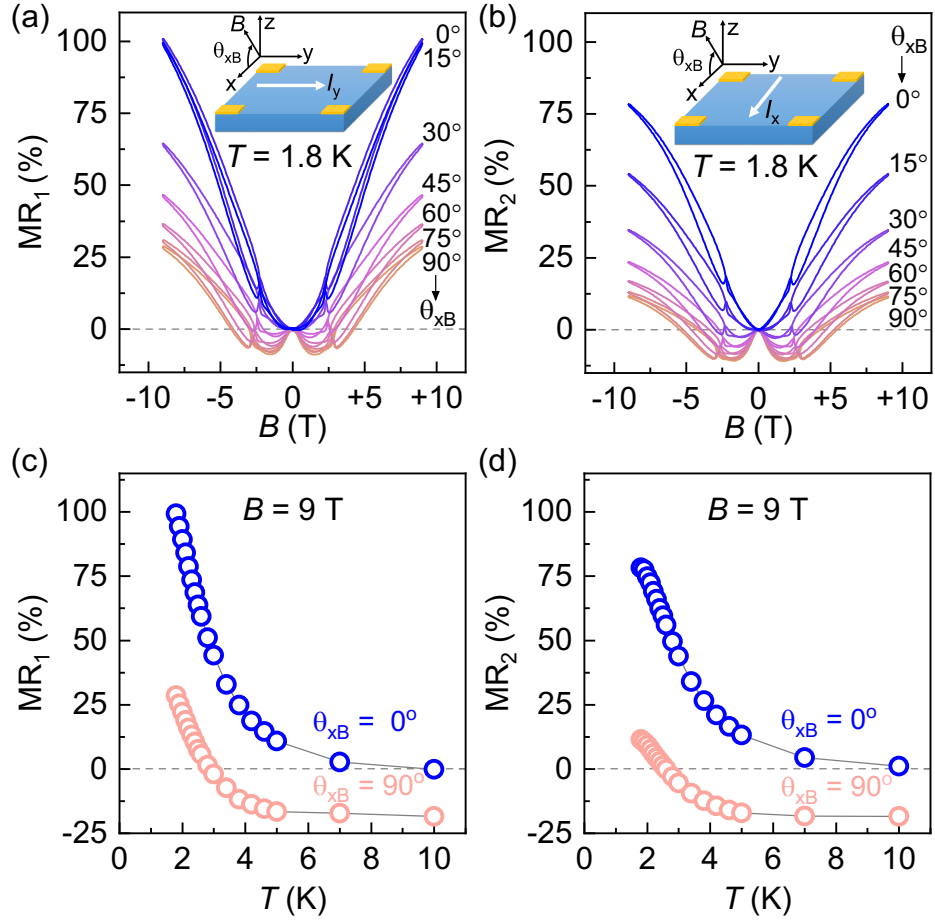

**Figure 3:** Angle-dependent MR at $T = 1.8$ K for two different configurations with (a) $I \perp x$ and (b) $I \parallel x$. Insets show the measurement configuration. $B$ is out-of-plane to the sample surface when $\theta_{xB} = 0°$ and in-plane when $\theta_{xB} = 90°$. (c, d) Temperature-dependent anisotropic MR at $B = 9$ T in each configuration for $\theta_{xB} = 0°$ and $\theta_{xB} = 90°$.

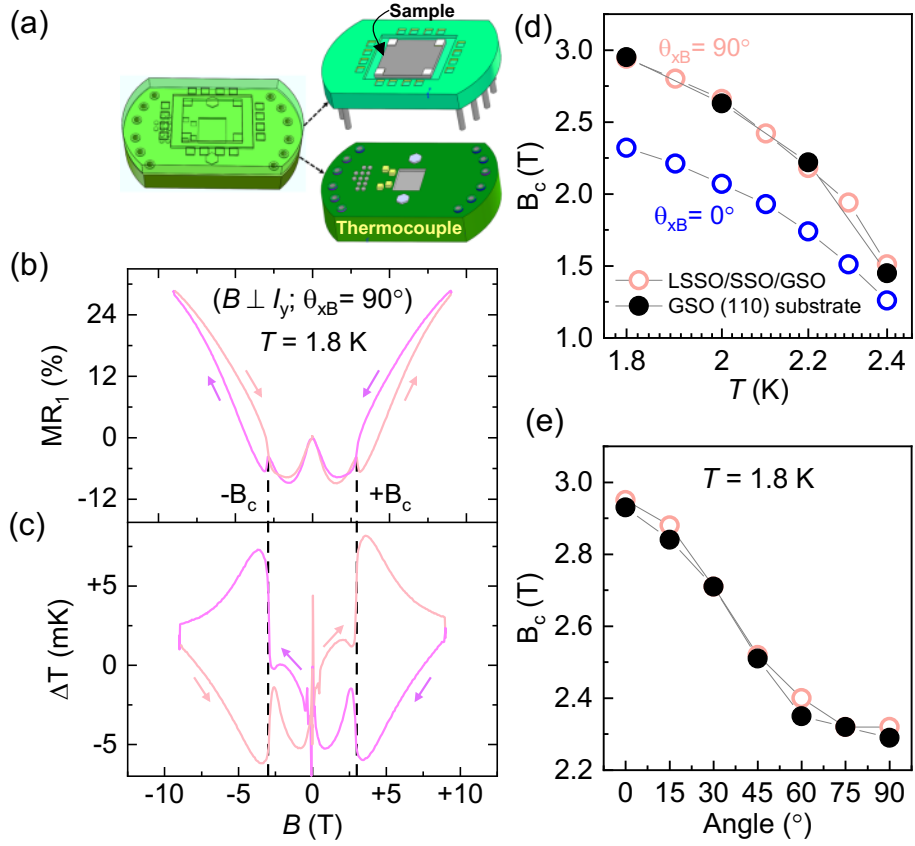

**Figure 4:** (a) Schematic of the measurement setup illustrating the position of the same and thermocouple. (b) Hysteretic MR at $T$ = 1.8 K for configuration 1 with the field out-of-plane ($\theta_{xB}$ = 90º). $B_c$ represents the critical field at which an anomaly in MR appears. (c) The evolution of the thermocouple temperature of the sample during the field sweep from 0 T → 9 T → -9 T → 0 T. (d) Temperature-dependent $B_c$ for 12 nm La:SSO/2 nm SSO/GSO and a bare GSO substrate for $\theta_{xB}$ = 90º and 0º. (e) Angle-dependent critical field at $T$ = 1.8 K for $B_c$ for a 12 nm La-doped SSO/2 nm undoped SSO/GSO and a bare GSO substrate.

18